\documentclass[a4paper,12pt]{article}
\usepackage{amsfonts}
\usepackage{amsmath}
\def\CA{{\cal A}}
\def\CB{{\cal B}}

\def\CF{{\cal F}}
\def\CM{{\cal M}} 

\def\CP{{\cal P}}

\def\BN{{\bf N}}

\def\Bmv{{\mbox{\boldmath$v$}}}

\def\real{\mathbb{R}}
\def\complex{\mathbb{C}}
\def\integer{\mathbb{Z}}

\def\be{\begin{equation}}
\def\ee{\end{equation}}
\def\ben{\begin{equation*}}
\def\een{\end{equation*}}
\def\bea{\begin{eqnarray}}
\def\eea{\end{eqnarray}}

\def\and{\quad{\rm and}\quad}

\def\half{{1\over2}}

\def\Ket#1{|#1 \rangle}
\def\id{{\bf1}}
\def\End{{\rm End}}

\newcommand{\eq}[1]{(\ref{#1})}

\def\nn{\nonumber}
\def\obar{\overline}
\def\beqa{\begin{eqnarray}} 
\def\eeqa{\end{eqnarray}} 
\def\beq{\begin{equation}} 
\def\eeq{\end{equation}}

\def\a{\alpha}        
\def\b{\beta}  
\def\d{\delta}

 \def\L{\Lambda} \def\la{\lambda}

\def\cA{{\cal A}}  
  \def\cF{{\cal F}}
  \def\cI{{\cal I}}
  
  \def\cO{{\cal O}}

\def\mg{\mathfrak{g}}

\def\R{{\mathbb R}}
\def\C{{\mathbb C}}

\def\one{\mbox{1 \kern-.59em {\rm l}}}

\def\({\left(}
\def\){\right)}
\def\tens{\otimes}
\def\rep{representation }
\def\reps{representations }

\begin{document} 
\begin{titlepage}
\null \hfill Preprint TU-715  \\
\null \hfill  LMU-TPW 01/04 \\
\null \hfill October 2004 \\[4ex]

\begin{center}
{\LARGE\bf Monopole Bundles over Fuzzy Complex Projective Spaces}\\[3em] 

{\large 
Ursula Carow-Watamura${}^a$\footnote{
e-mail: ursula@tuhep.phys.tohoku.ac.jp}, 
Harold Steinacker${}^b$\footnote{
e-mail: harold.steinacker@physik.uni-muenchen.de}
\  and \\
Satoshi Watamura${}^a$\footnote{e-mail: watamura@tuhep.phys.tohoku.ac.jp}
 \\ [2ex]
}
${}^a$Department of Physics \\
Graduate School of Science \\
Tohoku University \\
Aoba-ku, Sendai 980-8578, Japan \\ [2ex]
${}^b$Institut f\"ur theoretische Physik\\
Ludwig-Maximilians Universit\"at M\"unchen\\
Theresienstr. 37, D-80333 M\"unchen, Germany\\[2ex]

\begin{abstract}
We give a construction of the
monopole bundles over fuzzy complex projective 
spaces as projective modules.
The corresponding Chern classes  
are calculated. They reduce to the 
monopole charges in the $N\rightarrow \infty$ limit, where $N$
labels the representation of the fuzzy algebra.
\end{abstract}
\end{center}

\end{titlepage}

\section{Introduction}

Gauge theory on noncommutative spaces
has attracted considerable interest in recent years. In particular,
topologically nontrivial solutions such as instantons were found
on the noncommutative plane. They
are conveniently described in terms of projective modules over 
the algebra $\CA$ of functions on the noncommutative space.
Additional motivation is provided by
the appearance of noncommutative gauge theory as an effective theory of 
D-branes in string theory. This led to an interpretation of the
nontrivial solutions of gauge theory on noncommutative
flat space in terms of
nonperturbative configurations in the 
D-brane background, 
see e.g. \cite{Nekrasovtrieste}. 

On the other hand, 
gauge theory on non-flat noncommutative spaces is not very well understood. 
A particularly nice class of  such 
spaces are the so-called fuzzy spaces, the simplest example being
the fuzzy sphere \cite{Madore1991}.
Understanding field theory and in particular gauge theory on fuzzy
spaces is important for several reasons. First, the fuzzy spaces
provide a nice regularization of field theory, because they admit
only finitely many
degrees of freedom. This provides an alternative to
lattice regularization, with the advantage of preserving a large
symmetry group \cite{grosse-first,Grosse9902138}.
While much work has been done for the fuzzy sphere, the
higher-dimensional spaces such as fuzzy $\C P^n$ are largely
unexplored. 

The fuzzy spaces also arise in string theory.
For example, the fuzzy sphere and its $q$-deformed version appear 
as a D-brane in the $SU(2)$ WZW model, as discussed by several 
authors \cite{ARS9908040,FFFS}; 
see also \cite{IW0007141} and references therein.
In fact, many fuzzy spaces investigated so far can be considered 
as $D$-branes on group manifolds
\cite{qfuzzybranes}. 
Moreover, gauge theory on the fuzzy sphere appears
as an effective theory of the D-branes in $S^3$, in the $k\rightarrow\infty$ 
limit of the $SU(2)_k$ WZW model at level $k$. 
Furthermore, fuzzy spaces also arise  
as solutions of the IKKT matrix model \cite{iso,Kitazawa0207115}.
It is therefore natural to ask for a proper geometrical description
and interpretation of such a system, in particular of the 
topologically nontrivial configurations.

Topological aspects of field theory on the fuzzy 
sphere have first been discussed in \cite{Grosse9510083}. 
The formulation as projective modules
has been elaborated explicitly for the fuzzy sphere 
in \cite{Balachan9811169, Valtancoli0101189, GrosseMP0105033},
and an alternative approach using matrix models was given in
\cite{fuzzyYM}.

In ref. \cite{BaGoYd9911087,BaGoYd0006216} the authors have investigated 
the Dirac operators on the fuzzy sphere which have been proposed 
in \cite{Grosse9510083} and \cite{watamura9605003,watamura9710034} with 
respect to their differences, 
and their relation to topologically nontrivial configurations
and the fermion doubling problem have been studied in
\cite{BalImm0301242, Iso0209223}.

For physical applications, it is clearly desirable to consider spaces of
dimension 4 and higher.
The simplest higher-dimensional fuzzy spaces are fuzzy 
$\C P^2$ and $\C P^n$, which have been studied in 
\cite{Grosse9902138,Balachan0103023,Balachan0107099,Nair0203264}. 

In this paper, we consider fuzzy 
$\C P^2$ and $\C P^n$ in more detail, and we
present a simple formulation of monopole bundles 
on fuzzy $\C P^n$ using projective modules. 
We also introduce a suitable differential calculus, which allows to
compute the canonical connection and field strength explicitly.
The corresponding Chern classes are calculated. 
As in the case of the fuzzy sphere, the Chern numbers are 
integers only in the commutative limit.
For related work on a fuzzy four-sphere see \cite{valtan-04}.

The outline of this paper is as follows.
In section 2, the geometry of classical $\C P^n$ is formulated using 
two different approaches. The first is in terms of (co)adjoint orbits 
of $su(n+1)$, and the second using a generalized Hopf fibration. 
Both lead to a useful characterization in terms of $(n+1) \times (n+1)$
matrices satisfying a quadratic characteristic equation. 
For better readability we first present the case of $\C P^2$, and 
then the general case for $\C P^n$ separately.

In section 3, the fuzzy spaces $\C P^2_N$ and $\C P^n_N$ are discussed
form these two points of view, leading again to a 
quadratic characteristic equation for algebra-valued
$(n+1) \times (n+1)$ matrices.
This encodes the commutation relations of the coordinate algebra
in a compact way. In particular, the Hopf fibration is quantized 
in terms of a Fock space representation. 

In section 4, we give a construction of the projective modules 
corresponding to monopole bundles for fuzzy $\C P^n_N$. 
This is done using an explicit form of the projection operators. 
We also explain how a section of the constructed line 
bundles corresponds to a complex scalar field.
A similar construction for the classical case of $S^3\rightarrow S^2$ 
has been given by Landi in \cite{LandiMP9812004,LandiMP9905014}.

In section 5, a differential calculus is constructed, which in the
fuzzy case involves more degrees of freedom than in the classical
case. This is again typical for fuzzy spaces, and we explain in what
sense the classical calculus is recovered in the commutative limit.
This calculus is then used to
compute the field strength and Chern class for the monopole bundles.
We show that the usual (integer) Chern numbers are recovered in the 
limit of large $N$.

\section{The geometry of $\C P^n$}

We will discuss two descriptions of $\C P^n$ here. 
The first is in terms of 
(co)adjoint orbits of $su(n+1)$, and the second is 
based on the generalized Hopf fibration 
$U(1)  \to S^{2n+1} \to \C P^{n}$. 
Both are manifestly covariant under 
$SU(n+1)$, which is maintained in their quantization as 
fuzzy  $\C P^n$.

\subsection{Adjoint orbits}
\label{subsec:adorbits}

In general, an adjoint orbit of a 
(finite-dimensional) 
matrix Lie group
$G$ with Lie algebra $\mg$  is given in terms of 
some $t \in \mg$ as
\beq
\cO(t) = \{g t g^{-1}; \;\; g \in G\} \subset \mg.
\label{conj-classes}
\eeq
Then $\cO(t)$ can be viewed as a homogeneous space:
\beq
\cO(t) \cong G/K_t.
\label{coset}
\eeq
Here $K_t = \{g \in G:\; [g,t] = 0\}$ is the stabilizer of $t$,
which determines the nature of the space $\cO(t)$.
Any such conjugacy class is invariant under the adjoint action of
$G$. 
``Regular'' conjugacy classes are those with $K_t$ being 
the maximal torus, 
 and have maximal dimension $dim(\cO(t)) = dim(G) - rank(G)$. 
Here we are interested in degenerate orbits such as $\C P^2$ or $\C P^n$.
They correspond to degenerate $t$, and have dimension 
$dim(\cO(t)) = dim(G) - dim(K_t)$. 

A nice way to characterize the type of the orbit (for matrix Lie groups)
is through its characteristic equation $\chi(Y) =0$ for $Y \in \cO(t)$,
which is invariant under conjugation and therefore
depends only on the eigenvalues of $t$.

For example, 
in order to obtain $\C P^2 = SU(3)/(SU(2) \times U(1))$, 
we should choose the matrix 
$t\in \mg \cong \{Y \in Mat(3,\C);\;Y^\dagger = Y, tr(Y) =0\}$ 
with only two distinct eigenvalues.
A natural choice for an element of the adjoint orbit is hence
$t = diag(-1,-1,2)$,
so that the stabilizer is $K_t = SU(2) \times U(1)$.
The normalization of the matrix $t$ defines the scale of the 
resulting $\C P^2$, which is irrelevant for the 
discussion here\footnote{Notice that there is also the ``conjugated'' 
space with $t \cong diag(1,1,-2)$, which correspond to the 
charge conjugation of the $\C P^2$ defined here.}.

The characteristic equation for $\C P^2$ is therefore quadratic, and
has the form 
\beq
\chi(Y) = (Y+1)(Y-2) =0 \quad.
\label{char-1}
\eeq
This equation 
characterizes $\C P^2$ as submanifold in the 
embedding space $\R^8$.
We will see that an analogous characteristic equation 
holds for fuzzy $\C P^n$.

This construction of $\C P^2$ 
can also be understood as follows: 
The $3\times 3$ matrix
\beq
P = \frac 1{3}(Y + 1)\qquad \in Mat(3,\C)
\eeq
satisfies
\beq
P^2 = P, \qquad Tr(P) =1
\label{proj-class}
\eeq
as a consequence of \eq{char-1}, hence $P \in Mat(3,\C)$ 
is a projector of rank 1 and can be written as 
$$
P = |z^i\rangle\langle z^i| = (z^i)^\dagger (z^i)
$$
where $(z^i) = (z^1,z^2,z^3) \in \C^3$ is normalized as
$\langle z^i | z_i\rangle =1$.  
Such projectors are equivalent to rays in $\C^3$, which give the second
description of $\C P^2$ as $S^5/U(1)$. The adjoint action on $Y$ 
corresponds to the fundamental \rep on $\C^3$.

Similarly,
to obtain $\C P^n \cong SU(n+1) / (SU(n) \times U(1))$, we need a matrix
$t \in su(n+1)$ with 2 distinct eigenvalues and multiplicities $(n,1)$, 
hence a natural choice is
$t = diag(-1,-1,...,-1,n)$ up to normalization.
It satisfies the characteristic equation
\beq
\chi(Y) = (Y+1)(Y-n) =0
\label{char1}
\eeq
 Again, this can be understood by
considering the $(n+1)\times (n+1)$ matrix
\beq
P = \frac 1{n+1}(Y + 1) \qquad \in Mat(n+1,\C)
\eeq
which is also a projector of rank 1.
Hence $P \in Mat(n+1,\C)$ can be written as 
$$
P = |z^i\rangle\langle z^i| 
$$
where $\langle z^i|z_i\rangle =1$.  
This is the relation to the second description of $\C P^n$
as $S^{2n+1} / U(1)$.

\subsection{Global coordinates for $\C P^2$ and $\C P^n$}
\label{sec:coord-class}

For later purpose, we introduce coordinates
on $\C P^2$ and $\C P^n$. It is useful to choose 
an overcomplete set of global coordinates, 
which is easily generalized to the fuzzy case.
Let us first consider $\C P^2$, described as above by the matrix 
$Y = g^{-1} t g \in Mat(3,\C)$.
It is natural to write the matrix $Y$ in terms of the  
Gell-Mann matrices $\la_a$ of $su(3)$ as
\beq
Y = y_a \la_a.\label{cp2coordinate}
\eeq
The Gell-Mann matrices satisfy
\beqa
tr(\la_a \la_b) &=& 2 \d_{ab},\nn\\
\la_a \la_b &=& \frac 23 \d_{ab} + (i {f_{abc}} + {d_{abc}}) \la_c
\eeqa
where $f_{abc}$ are the totally antisymmetric structure constants, and
$d_{abc}$ the totally symmetric invariant tensors of $su(3)$.
The $\la_a$ are related to the generators $T_a$ of the Lie algebra via
\beq
\la_a = 2 \pi_{\Lambda_{(1)}}(T_a).
\eeq
Here $\pi_{\Lambda_{(1)}}$ denotes the fundamental \rep 
of $su(3)$ with highest weight $\Lambda_{(1)}$. 
The characteristic equation 
\beq
Y^2 =  Y +  2\quad
\label{char-class}
\eeq
written in terms of the coordinates $y_a$ in eq.(\ref{cp2coordinate})
takes the form 
\beq
g_{ab}\; y_a y_b    = 3, \qquad
d_{abc}\; y_a  y_b =  y_c. \label{def3}
\eeq
It is clear from the above construction 
that this set of relations indeed characterizes
the appropriate adjoint orbit in $su(3)$.

For $\C P^n$, we consider the generalized 
Gell-Mann matrices of $su(n+1)$ which are defined by
\beq 
\la_a = 2\pi_{\Lambda_{(1)}}(T_a)
\eeq
where $T_a$ are the generators of $su(n+1)$.
They satisfy 
\be
\la_a \la_b = \frac 2{n+1} \d_{ab}+(i {f_{abc}} + {d_{abc}}) \la_c.
\ee
Then the characteristic equation 
\beq
Y^2 = (n-1) Y + n
\label{char-class-n}
\eeq
using the expansion $Y = y_a \la_a = g^{-1} t g$
takes the form 
\beq
g_{ab}\; y_a y_b    = \frac{n(n+1)}2, \qquad
d_{abc}\; y_a  y_b = (n-1)\; y_c. \label{def3-n}
\eeq

\paragraph{Some geometry.}

Notice that the symmetry group $SU(3)$ contains both ``rotations'' as
well as ``translations''. The generators $J_a$
act on a point $Y = y_a \la_a \in \C P^2$ as
\beq
J_a Y = \frac 12 [\la_a,Y] = \frac 12 y_b [\la_a,\la_b] 
  = i {f_{abc}}\; y_b \la_c.
\eeq
In terms of the coordinate functions on the embedding space $\R^8$,
this can be realized as differential operator
\beq
J_a = \frac i2 f_{abc}(y_b \partial_c - y_c \partial_b).
\eeq
Now we can identify the rotations in $SU(3)$: 
Consider the ``south pole'' on $\C P^2$, for $Y = diag(-1,-1,2) =-r\la^8$
for the radius $r = \sqrt{3}$,
hence $y_a = -r \d_{a,8}$. The rotation subgroup is generated by 
its stabilizer algebra $su(2) \times u(1)$, which is generated by the elements
\beq
\{\la_1,\la_2,\la_3,\la_8\}
\eeq
of $su(3)$ (using the standard conventions). It is a
subgroup of the Euclidean rotation group $SO(4) = SU(2)_L \times SU(2)_R$.
The remaining generators
\beq
\{\la_4,\la_5,\la_6,\la_7\}
\label{tang-indices}
\eeq
change the position of $Y = -r\la_8 \in \C P^2$, hence they correspond to 
``translations''. All this generalizes to $\C P^n$ in an obvious way.

There is another interesting subspace of $\C P^2$: the ``north sphere''.
This is a nontrivial cycle of $\C P^2$ which will be useful later.
Consider again the parametrization of $\C P^2$ in terms of 
$3 \times 3$ matrices $Y = U^{-1} diag(-1,-1,2) U$
introduced in section \ref{subsec:adorbits}. Using a suitable 
$U \in SU(3)$, we can put it into the form
\beq
Y = \left(\begin{array}{ccc} \frac 12 + y_i \sigma^i & \vline & 0\\
                             \hline 
                             0 & \vline & -1 \end{array}\right).
\label{sphere-matrix}
\eeq
This is the subspace of $\C P^2$ with maximal value of 
$y_8 = \frac 12 r$, where
$y_{4,5,6,7} =0$ and 
$2 r^2 = Tr Y^2 = 2 \sum_{a=1,2,3,8} y_a^2$. 
It follows that $y_1^2+y_2^2+y_3^2 = \frac 34 r^2$, which is a sphere
of radius $\frac{\sqrt{3}}2  r$. 

A similar sphere can be found for all $\C P^n$: consider again matrices of
the form \eq{sphere-matrix} with $n-1$ entries $-1$ in the lower right
block. Then the upper left block has the form 
$\frac {n-1}2 + y_i \sigma^i$, which has eigenvalues $(-1,n)$ provided
$y_1^2+y_2^2+y_3^2 = \frac{n+1}{2n}\; r^2$, hence it is a sphere
of radius $\sqrt{\frac{n+1}{2n}}\; r$. Here $r$ is the radius of $\C P^n$.
One can now choose a Gell-Mann basis of $su(n+1)$ which contain
the above $\sigma^i$, so that all other $dy_a$ for $a \neq 1,2,3$ 
vanish on this sphere.
This implies that these are non-trivial cycles, see section 
\ref{sec:chern-int}.

\subsection{Harmonic analysis}

Fuzzy $\C P^n$ is defined as a particular (finite) 
noncommutative algebra which 
is covariant under $SU(3)$, and 
is interpreted as quantization of the algebra of functions of $\C P^n$.
To justify the construction, 
we first have to understand the space of harmonics 
on $\C P^n$, and then compare it with the noncommutative case.
This can be done using the Hopf fibration $U(1) \to S^{2n+1} \to \C
P^{n}$.

First, consider the space of equivariant functions over $S^{2n+1}$.
Define the $U(1)$ action on $\C^{n+1}$ as
\be
\omega\circ (z^i, \obar{z}_i)
=(z^i\omega, \obar{z}_i \overline{\omega}) 
\label{U1-action}
\ee
where $\omega\in U(1)$. With this $U(1)$ action, 
we define the equivariant functions $C(\kappa, \C^{n+1})$
of $z^i ,\obar{z}_j$ by\footnote{ 
If we follow strictly the Hopf fibration, we should first impose the
$SU(n+1)$ invariant condition $\sum z^i \obar{z}_i=1$ and consider
functions over $S^{2n+1}$. 
However we impose this constraint later, which is more appropriate
in the fuzzy case.}
\beqa
C(\kappa, \C^{n+1}) &=& \{f\in Pol(z^i,\obar{z}_j), i,j=1,...,n+1 
  \and \omega \circ f=f\omega^\kappa, \} \nn\\
 &=& \mathop{\oplus}_p\{ Pol_{p,q}(z,\obar{z}), \quad p-q=\kappa\}
\eeqa
where  $Pol_{p,q}(z,\obar{z})$ denotes the polynomial functions 
of degree $(p,q)$ in the coordinates $z^i$ resp. $\obar{z}_j$
on $\C^{n+1}$.
As a representation of $su(n+1)$, it has the structure
$$
Pol_{p,q}(z,\obar{z}) \cong V_{(p,0,...,0)} \tens V_{(0,...,0,q)}
$$
where $V_{(d_1,...,d_n)}$ denotes the highest weight irrep 
(=irreducible representation) with Dynkin indices $(d_1,...,d_n)$. 

In order to identify these with the functions on $S^{2n+1}$, 
we impose the condition 
$r^2 = \sum_i z^i\obar{z}_i$, which defines the equivariant 
functions $C(\kappa, S^{2n+1})$. 
By construction, it is clear that 
 $C(0, S^{2n+1})$ is isomorphic to the space of 
functions $C(\C P^n)$ on $\C P^n$.
Since $V_{(p,0,...,0)} \tens V_{(0,...,0,p)}=\mathop{\oplus}_{n=0}^p
V_{(n,0,...,0,n)}$ and taking the radius $r$ into account,
it follows that $C(0, S^{2n+1})$ decomposes under $su(n+1)$ as
\beqa
C(\C P^n)\cong C(0, S^{2n+1})&=&\mathop{\oplus}_{p=0}^\infty V_{(p,0,...,0,p)}.
\label{CP2-harmonics}
\eeqa
Similarly,  
$C(\kappa, S^{2n+1})$ can be identified with 
the space of sections $\Gamma_\kappa(\C P^n)$ of the line bundle on 
$\C P^n$ with monopole number $\kappa$.
Moreover, there is a natural multiplication of two equivariant polynomials  
$C(\kappa,S^{2n+1})$ and $C(\kappa',S^{2n+1})$ such that
\be
C(\kappa,S^{2n+1})\times C(\kappa',S^{2n+1})
\longrightarrow C(\kappa+\kappa',S^{2n+1}).
\ee
Therefore $\Gamma_\kappa(\C P^n) \cong C(\kappa,S^{2n+1})$ 
is a module over $C(\C P^n)$. 
The decomposition under $su(n+1)$ is similar to (\ref{CP2-harmonics}):
\beqa
\Gamma_\kappa(\C P^n) \cong C(\kappa,S^{2n+1}) &=& 
   \mathop{\oplus}_{n=0}^\infty V_{(n,0,...,0,n+\kappa)}
\label{CP2-moduleharmonics}
\eeqa
see also \cite{Grosse9902138}.
We will recover this structure of harmonics in the fuzzy case, 
up to some cutoff.

\section{Fuzzy Complex Projective Spaces $\C P^n_N$.}

In general, 
(co)adjoint orbits \eq{conj-classes} on $G$ can be quantized in terms 
of a simple matrix algebra
$End_\C(V_N)$, where $V_N$ are suitable representations of $G$. 
The appropriate representations 
$V_N$ can be identified 
by matching the spaces of harmonics (i.e. using harmonic analysis),
see \cite{qfuzzybranes} for the general case. 
Fuzzy $\C P^2$ has been introduced in \cite{Grosse9902138,Balachan0103023},
and fuzzy $\C P^n$ in \cite{Balachan0107099}.

\subsection{$\C P^2_N$}
Again we first consider $\C P^2$. To identify the correct \reps
$V_N$ of $su(3)$, we must match the space of harmonics \eq{CP2-harmonics} 
with the decomposition of 
$End_\C(V_N)$ under the adjoint, which is 
\beq
End(V_N) =  V_N \tens V_N^* = \mathop{\oplus}_\la n_\la V_{\la}
\eeq 
for certain multiplicities $n_\la$. Here $\lambda$ denotes the highest weight.
It is easy to see that 
for both $V_N= V_{(N,0)}$ and
$V_N'= V_{(0,N)} = V_N^*$, we have
\beq
V_N \tens V_N^* = V_{(N,0)} \tens V_{(0,N) }
   \cong \mathop{\oplus}_{p=0}^{N}\; V_{(p,p)}
\label{decomposition}
\eeq
which matches \eq{CP2-harmonics} up to a cutoff. 
Therefore we define fuzzy $\C P^2$ as 
\beq
\C P^2_N:= End_\C(V_N) = Mat(D_N,\C)
\eeq
for $V_N = V_{(N,0)}$,\footnote{
Alternatively, $\C P_N^{2*}$ for $V_N' = V_{(0,N)}$, which is equivalent as algebra} 
where
\beq
D_N=dim(V_N) = (N+1)(N+2)/2.
\eeq
Under the (adjoint) action of $su(3)$,
it decomposes into the harmonics \eq{decomposition} 
$\mathop{\oplus}_{p=0}^{N}\; V_{(p,p)}$, 
cp. \cite{Grosse9902138}. 
Comparing with (\ref{CP2-harmonics}), these harmonics are in
one to one correspondence with the harmonics on classical 
$\C P^2$ up to the cutoff 
at $p=N$. The remarkable point is that
 this finite space of
harmonics closes under the matrix multiplication in $\C P^2_N$.
Hence by construction, fields on $\C P^2$ can be approximated by
$\C P^2_N = Mat(D_N,\C)$,
therefore field theory on fuzzy 
$\C P^2_N$ should be a good
regularization for field theory on $\C P^2$.

To make the correspondence with classical $\C P^2$ 
more explicit, we consider the 
$3D_N \times 3D_N$ matrix 
\beq
X = \sum_a \xi_a \la_a 
\label{X-intro}
\eeq
where $\la_a$ are the Gell-Mann matrices as before, and  
\beq
\xi_a = \pi_{V_N}(T_a)  \qquad \in \C P^2_N. 
\eeq
denotes the representation of $T_a \in su(3)$ on $V_N$.
The coordinate functions $x_a= (x_1, ... x_8)$
on fuzzy $\C P^2$ are defined by
\beq
x_a = \Lambda_N \xi_a  \qquad \in \C P^2_N \quad.
\eeq
They are operators acting on $V_N$. 
Here $\Lambda_N$ is a scaling parameter which will be fixed below.
By construction,
the $x_a$ transform in the adjoint under $su(3)$, just like the classical 
coordinate functions $y_a$ introduced in Section \ref{sec:coord-class}.

To find the relations among these generators $x_a$, we can use the 
characteristic equation \eq{char-X-n} of $X$ given in appendix:
\beqa
X^2 &=& \xi_a \xi_b \;
  (\frac 23 \d_{ab} + (i {f_{abc}} + {d_{abc}}) \la_c)  \nn\\
 &=& \frac 23 (\frac 13 N^2 +N)  +(\frac{N}3 -1) X.
\eeqa
Using  $f_{abc} f_{dbc} = 3 \d_{ad}$, we obtain

\beqa
i f_{abc}\; \xi_a  \xi_b &=& -\frac 32 \; \xi_c, 
    \quad [\xi_a, \xi_b] = i f_{abc} \xi_c  \label{defxi1}\\
g_{ab}\; \xi_a \xi_b    &=&  (\frac 13 N^2 +N), \label{defxi2} \\
d_{abc}\; \xi_a  \xi_b &=& (\frac{N}3 + \frac 12)\; \xi_c.   \label{defxi3}
\eeqa
Hence taking the scale parameter $\Lambda_N$ to be
\beq
\Lambda_N = \frac 1{\sqrt{\frac 13 N^2 +N}}\quad,
\eeq
we find the defining relation of the algebra $\C P^2_N$:
\beqa
[x_a, x_b] &=&  \frac i{\sqrt{\frac 13 N^2 +N}}f_{abc}\; x_c,  \label{defz1}\\
g_{ab}\; x_a x_b    &=& 1, \label{defz2} \\
d_{abc}\; x_a x_b &=& \frac{\frac{N}3 +\frac 12}{\sqrt{\frac 13 N^2 +N}}\; 
   x_c   \label{defz3}
\eeqa
in agreement with Balachandran et al. \cite{Balachan0103023}.
 
Let us verify that $\C P^2_N$  admits an approximate ``south pole''
at $x_8 \approx -1$. It corresponds to the lowest weight state of 
$V_{(N,0)}$ which has eigenvalue $\xi_8 = -\frac{2N}{\sqrt{3}}$, hence
$x_8 = -\frac N{\sqrt{ N^2 +3N}} \approx -1$. For $\C P^{2*}_N$, 
one would obtain an approximate north pole, but no south pole.

The ``angular momentum'' operators (generators of $SU(3)$) now 
become inner derivations:
\beq
J_a f(x) =  [\xi_a,f],
\label{angular-momentum}
\eeq
and $J_a x_b = [\xi_a,x_b] = i {f_{abc}} x_c$,
as classically. 
Recall that the symmetry group $SU(3)$ contains both ``rotations'' as
well as ``translations''.

The integral on $\C P^2_N$ is given by the suitably normalized trace,
\beq
\int f(x) = \frac 1{D_N}\; Tr(f)= \frac 2{(N+1)(N+2)}\; Tr(f) 
\eeq
which is clearly invariant under $SU(3)$.

\subsection{$\C P^n_N$}

The construction for fuzzy  $\C P^n$ is entirely analogous to fuzzy
$\C P^2$.
To identify the correct \reps
$V_N$ of $su(n+1)$, we must match the space of harmonics \eq{CP2-harmonics} 
with the decomposition of 
$End(V_N) =  V_N \tens V_N^* = \mathop{\oplus}_\la n_\la V_{\la}$
for certain multiplicities $n_\la$. 
Similar to the case of $su(3)$, it is easy to show that 
\beq
V_N \tens V_N^*\cong \mathop{\oplus}_{p=0}^{N}\; V_{(p,0,....,0,p)}\label{decomp_CPn}
\eeq
where 
$$
V_N := V_{(N,0,...,0)}\quad.
$$
The representations appearing in the r.h.s. of eq.(\ref{decomp_CPn}) 
match \eq{CP2-harmonics} up to the cutoff $N$. 
We therefore define the algebra of the fuzzy projective space by
\beq
\C P^n_N := End_\C(V_N) = Mat(D_N,\C)
\eeq
where
\beq
D_N = dim(V_N) = {(N+n)!\over n!\,N!}
\eeq
from Weyl's dimension formula.
The fuzzy coordinates and their commutation relations are obtained 
again by considering the 
$(n+1)D_N \times (n+1)D_N$ matrix 
\beq
X = \sum_a \xi_a \la_a \quad,
\eeq
where $\la_a$ are the Gell-Mann matrices of $su(n+1)$ and  
\beq
\xi_a = \pi_{V_N}(T_a)  \qquad \in  \C P^n_N. 
\eeq
The coordinate functions $x_a$ for $a=1,...,n^2+2n$
on fuzzy $\C P^n$ are defined by
\beq
x_a = \Lambda_N \xi_a  \qquad \in \C P^n_N.\label{rescaling}
\eeq
$\Lambda_N$ is a scaling parameter which will be fixed below.
By construction,
the $x_a$ transform in the adjoint under $su(n+1)$, just like the classical 
coordinate functions $y_a$ introduced in Section \ref{sec:coord-class}.
Using the 
characteristic equation \eq{char-X} of $X$,
\be
(X-\frac{nN}{n+1})(X+\frac{N}{n+1}+1)=0\label{characteristiceq}
\ee
we have
\beqa
X^2 &=& \xi_a \xi_b \;
  (\frac 2{n+1} \d_{ab} + (i {f_{abc}} + {d_{abc}}) \la_c),  \nn\\
 &=& \frac n{n+1} (\frac 1{n+1} N^2 +N) +(\frac{N(n-1)}{n+1} -1) X.
\eeqa
Using  $f_{abc} f_{dbc} = (n+1) \d_{ad}$, 
we obtain
\beqa
i f_{abc}\; \xi_a  \xi_b &=& -\frac{n+1}2 \; \xi_c, 
    \quad [\xi_a, \xi_b] = i f_{abc} \xi_c,  \label{defxi1-n}\\
g_{ab}\; \xi_a \xi_b    &=& \frac n2 (\frac 1{n+1} N^2 +N), \label{defxi2-n} \\
d_{abc}\; \xi_a  \xi_b &=&(n-1)(\frac{N}{n+1}+\frac 12)\;\xi_c.\label{defxi3-n}
\eeqa
Hence for 
\beq
\Lambda_N = \frac 1{\sqrt{\frac{n}{2(n+1)} N^2 +\frac n2 N}}
\eeq
we find
\beqa
[x_a, x_b] &=&  i \Lambda_N f_{abc}\; x_c,  \label{defz1-n}\\
g_{ab}\; x_a x_b    &=& 1, \label{defz2-n} \\
d_{abc}\; x_a x_b 
&=& (n-1)(\frac{N}{n+1} +\frac 12) \Lambda_N\;x_c. \label{defz3-n}
\eeqa
For large $N$, this reduces to \eq{def3-n}.
Again, $\C P^n_N$  admits an approximate ``south pole''
which corresponds to the lowest weight state of $V_N$.
The symmetry group $SU(n+1)$ acts by inner derivation as for $\C P^2$, and the
integral is given by the suitably normalized trace over $V_N$.

\subsection{Representation on a Fock space}

In order to introduce nontrivial line bundles, 
it is useful to quantize directly the fibration 
$U(1)  \to S^{2n+1} \to \C P^{n}$. In this section we will
first introduce noncommutative $\C^{n+1}$, in terms of 
operators $a^i, a^+_i$ ($i=1,...,n+1$) which are quantizations of
the coordinate functions 
$z^i, \obar{z}_i$ of $\C^{n+1} \supset S^{2n+1}$. Then
fuzzy $\C P^n$ will be obtained as a subalgebra of this noncommutative
$\C^{n+1}$. 
This has been used first in \cite{Balachan0103023}.
The equivalence to the definition given previously will be manifest.

The generators $a^+_i, a^i$ of noncommutative $\C^{n+1}$ 
are creation- resp. annihilation operators
which transform as $V_{(1,0...,0)}$ resp. $V_{(1,0,...,0)}^* = V_{(0,..,0,1)}$, and satisfy
the canonical commutation relations
\beq
[a^{i} , a^{j} ] = [a_i^+ , {a_j^+} ] = 0 \ ,\ \
[a^{i} , a_j^+ ] = \delta^i_{j}.
\label{A_cr}
\eeq
The resulting algebra will be denoted as $\C^{n+1}_\theta$.

As in the previous section, we consider the $U(1)$ defined by
\be
\omega\circ (a^i, a^\dagger_i)
=(a^i\omega, a_i^+ \overline{\omega}) 
\ee
where $\omega\in\complex$ and $|\omega|=1$. With this $U(1)$ action, 
the equivariant operators $C(\kappa, \complex^{n+1}_\theta)$ are defined by
\be
C(\kappa, \complex^{n+1}_\theta)=\{f|f\in Pol(a^i,a_i^+), i=1,...,n+1 \and  \omega \circ f=f\omega^\kappa \}.
\ee
This means that $\kappa$ counts the difference of the number of creation 
and annihilation operators, and thus an element
 $f\in C(\kappa, \complex^{n+1}_\theta)$ satisfies
\be
 [\BN, f]=\kappa f,
\ee
where $\BN$ is the number operator: $\BN=\sum_{i=1}^{n+1}
a_i^+ a^i$.

As in the commutative case, there is a natural multiplication 
of the equivariant operators 
$C(\kappa,\complex^{n+1}_\theta)$ and
$C(\kappa',\complex^{n+1}_\theta)$ 
such that
\be
C(\kappa,\complex^{n+1}_\theta)\times C(\kappa',\complex^{n+1}_\theta)
\longrightarrow C(\kappa+\kappa',\complex^{n+1}_\theta).
\ee
Hence the equivariant operators
$C(\kappa,\complex^{n+1}_\theta)$ can be 
interpreted as $C(0, \complex^{n+1}_\theta)$-module.

Consider the following generators of $C(0, \complex^{n+1}_\theta)$:
\beq
\tilde\xi_a = a^+_i {(T_a)^i}_{j} a^j, \qquad T_a = \frac 12  \la_a.
\eeq
By construction, they transform in the adjoint of $su(n+1)$, and
satisfy the relations  
\be
[\tilde\xi_a,\tilde\xi_b]=if_{abc}\tilde\xi_c
\ee
and 
\be
\sum_{a=1,...,n^2+2n}\tilde \xi_a\tilde \xi_a
={n\over2(n+1)}\BN(\BN+n+1)\quad.
\ee

To obtain fuzzy $\C P^n_N$, we must ``fix the radius'' in 
$\complex^{n+1}_\theta$, i.e. choose a 
Fock space representation with fixed particle number $N$.
As usual, the Fock space is defined  by 
acting with the creation operators $a_i^+$ on the vacuum state $\Ket{0}$,
which satisfies $a_i\Ket{0} =0$.
The $N$-particle subspace $\cF_N$ is obtained
by acting with $N$ creation operators on the vacuum, with basis 
\be
\Ket{\vec m}=\sqrt{m_1!m_2!\cdots m_{n+1}!\over N!}(a_1^+)^{m_1}(a_2^+)^{m_2}\cdots (a_{n+1}^+)^{m_{n+1}}
\Ket{0}\label{fockspacebase}
\ee
where the label $\vec m$ is a set of positive integers $m_i$, 
$\vec m=(m_1,...,m_{n+1})$ satisfying $\sum_i m_i=N$. 

Hence if acting on the $N$-particle subspace $\cF_N$, we recover
precisely the relations \eq{defxi1-n} and \eq{defxi2-n}.
Next, we verify that the operators $\tilde\xi_a$ also satisfy
\eq{defxi3-n} if acting on $\cF_N$. 
In order to see this, we prove the characteristic 
equation \eq{characteristiceq} as follows: 
Using the Fierz identity for the generators $T^a$ and the definition  
of the generators $\tilde\xi$, we obtain 
\be
a^+_ja^i
=2\sum T^a{}^i_{j}\tilde\xi^a
+{1\over n+1} \delta^i_{j}\BN
=a^ia^+_j-\delta^i_{j}.
\label{fierzTa}
\ee
On the other hand, it is straightforward to confirm that the operator $P$:
\be
P^i{}_j={1\over\BN+1}a^i a^+_j
\ee
is a projection operator, i.e. $P^2=P$.
Defining the matrix $\tilde X^i_j =2\sum T^a{}^i_{j}\tilde \xi^a$ and 
using (\ref{fierzTa}), the operator $P$ can be written as 
\be
P={\tilde X+1+{\BN\over n+1}\over
\BN+1}\quad.
\ee
In terms of $\tilde X$, the relation $P^2=P$ gives
\be
0=P(P-1)={1\over (\BN+1)^2}(\tilde X+1+{\BN\over n+1})(\tilde X-\BN+{\BN\over n+1})\quad.
\ee
This is exactly the characteristic equation (\ref{characteristiceq}) 
for $X$, if the number 
operator $\BN$ is replaced by the number $N$.  
Hence the relations \eq{defxi1-n}, \eq{defxi2-n} and \eq{defxi3-n} 
of fuzzy $\C P^n$  hold on the Fock space
representation $\CF_N$ of $C(0, \C^{n+1}_\theta)$.

Therefore any $f\in C(0,\C^{n+1}_\theta)$ defines a map
\be 
f : \CF_N\longrightarrow \CF_N.
\ee
Now observe that the Fock space $\CF_N$ is precisely the irreducible
representation $V_N=V_{(N,0..,0)}$ of $su(n+1)$.
Since the generators $\tilde\xi_a$ act on $\CF_N$ and generate the 
$su(n+1)$ algebra, they generate $End_\C(\CF_N)$, and the
equivalence of this definition of $\C P^n_N$ with the one given 
in the previous section
in terms of $Mat(D_N,\C)$ (or more precisely $End_\C(V_N)$) is 
manifest\footnote{The 
conjugated version $\C P^{2*}$ would be obtained using the conjugated
$\la_a$ matrices $\tilde \la_a$, and 
$b^{+i}, b_j$ transforming in the dual representations.}.

\section{Projective modules}

{}From the noncommutative geometry point of view, the algebraic 
object corresponding to a vector bundle is the projective module 
of finite type. This equivalence of 
vector bundles and projective modules is based on the Serre-Swan Theorem, and 
thus in the following discussion, the projective modules are 
the relevant objects to deal with. 
 
A projective $\CA$-module can be constructed from the free module 
$\CA^p$ together 
with a projection operator $\CP$, which is an element of 
$Mat(p,\cA)$, the space of $p \times p$ matrices with 
elements in the base algebra $\CA$.

We will consider the noncommutative analogue of the monopole bundles, 
i.e. the 
$U(1)$ bundles over the fuzzy $\C P^n$. For this purpose, we construct a
rank $1$ projection operator which determines the module associated with the 
complex rank $1$ vector bundle. 
The advantage of this formulation compared to 
\cite{Grosse9902138} is that 
it also provides a canonical connection, 
which can be used e.g. to calculate Chern numbers.

We follow the approach taken in \cite{LandiMP9812004} here.
In general, a rank $1$ projection operator 
$\CP\in \End(\CA^{p})$ in the space of 
$p\times p$ matrices 
can be constructed by using an $p$-component 
vector defined as follows: 
\be
\Bmv=(v_\mu),\quad \mu=1,...,p
\ee
where $v_\mu $ is an element of a 
left-$\CB$-right-$\CA$ bimodule $\CM$.
Here $\CB$ is also an algebra, but not necessarily equivalent to $\CA$.
The only condition needed is the normalization condition
\be
\Bmv^\dagger \Bmv=\sum_\mu v_\mu ^\dagger v_\mu =\id_\CB
\ee
where $\id_\CB$ denotes the identity of the algebra $\CB$.
Using this vector $\Bmv$, we define the projection operator as
\be
\CP=\Bmv  \Bmv^\dagger.  \label{projection}
\ee
It is apparent that
\be
\CP\CP=\Bmv  (\Bmv^\dagger \Bmv) \Bmv^\dagger 
=\Bmv \id_\CB \Bmv^\dagger =\CP
\ee
by the normalization condition. 

Note that in order to define the projective module, 
the matrix elements of the projection operator $\CP$
must be elements of the algebra $\CA$. 
However, this does not mean that
each element of the vector $\Bmv$ is also an element of $\CA$.
Recall that a similar situation occurs 
in the construction of the so-called localized instanton in $\real^4_\theta$
using the ADHM 
construction \cite{Schwarz9802068,Furuuchi9912047,Nekrasovtrieste}.

When we construct the vector $\Bmv$ below, 
we take the elements to be in $C(\kappa,\complex^{n+1}_\theta)$.
Then the matrix elements of the projection operator $\CP_\kappa$ defined as 
in (\ref{projection}) are indeed elements of 
$C(0,\complex^{n+1}_\theta)$, which if acting on $\cF_N$ 
is just $\C P^n_N$, as it should.

To define the vector $\Bmv$, 
we should distinguish two cases depending on the sign of the integer $\kappa$:

\begin{enumerate}
\item For $0<\kappa$:
\be
v(\vec j) =(a^1)^{j_1}(a^2)^{j_2}\cdots (a^{n+1})^{j_{n+1}} c_+(\vec
j) 
\label{basevector+}
\ee
where the dimension of the vector $\Bmv$ is $D_\kappa={(n+\kappa)!\over n!\kappa!}$, 
${\vec j}=(j_1,...,j_n)$ where 
$j_i\geq0$ are integers with $\sum j_i=\kappa$.  
The normalization factor is
\be
(c_+(\vec j))^2={\kappa!\over j_1!j_2!\cdots j_{n+1}! \BN(\BN-1)\cdots(\BN-\kappa+1)}
\ee
\item For $-N<\kappa<0$:
\be
v_\mu =(a_1^+)^{j_1}(a^+_2)^{j_2}\cdots(a^+_{n+1})^{j_{n+1}} c_-(\vec j) 
\label{basevector-}
\ee
where 
the dimension is $D_\kappa={(n+|\kappa|)!\over n!\,|\kappa|!}$, and
the normalization is
\be
(c_-(\vec j))^2={|\kappa|! \over j_1!j_2!\cdots j_{n+1}! (\BN+n+1)(\BN+n+2)\cdots(\BN+n+|\kappa|)}
\ee
\end{enumerate}
It is easy to verify in both cases that $\Bmv^\dagger \Bmv =\id_\CB$.

One might worry that the denominator of the normalization factor 
$c_+(\vec j)$ can become $0$ for large $\kappa>0$.  
However, when constructing the projection operator  
$\CP_\kappa=\Bmv\Bmv^\dagger$ and specifying the representation space 
to be $\cF_N$, 
the number operator $\BN$ in $c_+$ is replaced  
by the value $N+\kappa$. 
Therefore for $\kappa>0$  
all expressions are well-defined.

On the other hand, for $\kappa<0$, there 
is a limit for the admissible values of  
$\kappa$. The reason is that 
$\Bmv^\dagger$ contains $|\kappa|$ annihilation operators, hence 
$$
\CP_\kappa=\Bmv\Bmv^\dagger
$$ 
acting on $\CF_N$ is ill-defined
if $|\kappa|>N$.
Therefore we must impose the bound $\kappa+N \geq 0$ in the case $\kappa<0$.

\subsection{Scalar fields, or sections in line bundles}

Now we can consider the projective module 
$\Gamma_\kappa(\C P^n_N)=\CP_\kappa
(\C P^n_N)^{D_{|\kappa|}}$.
An element $\xi$ of the module $\Gamma_\kappa(\C P^n_N)$ is thus a 
$D_{|\kappa|}$ dimensional vector 
$\xi=\{\xi_\mu\}$, the components of which are $\xi_\mu\in\C P^n_N$. 
This is a section of the line bundle, corresponding to a complex scalar
field in $U(1)$ gauge theory. 
However, the complex scalar field in $U(1)$ gauge theory 
has a single component in the conventional field theory formulation. 
The relation between these formulations will be explained next. 

Assume that $\kappa+N\geq 0$.
The single-component scalar field, i.e. 
section of the monopole bundle, is given by 
\be
\hat{\xi}=\Bmv^\dagger\xi=\sum_\mu v^\dagger_\mu \xi_\mu.\label{scalarfield}
\ee
On the other hand, 
we can act with  $\hat\xi$ 
on an element $\psi=\sum_m f_m\Ket{m}\in \CF_N$, with the result
\be
\hat\xi \psi=(\sum_\mu v_\mu ^\dagger \xi_\mu)(\sum_m f_m \Ket{m})
 = \sum_{{p_i\in\integer_+,\atop p_1+...+p_{n+1}
=N+\kappa} } f'_{p_1\cdots p_{n+1}} 
(a^+_1)^{p_1}\cdots 
(a_{n+1}^+)^{p_{n+1}}\Ket{0},
\ee
i.e.
\be
\hat\xi\psi \in \CF_{N+\kappa}. 
\label{xipsi}
\ee
Thus we can identify an element $\hat\xi\in C(\kappa,\C P^n_N)$ with a map
\be
\hat\xi\in Hom_\C(\cF_N,\cF_{N+\kappa}): \CF_N\longrightarrow \CF_{N+\kappa},
\ee
in agreement with \cite{Grosse9902138}.
In other words, we can identify the scalar field $\hat\xi$ on $\C P_N^n$
with monopole charge $\kappa$ with a 
$D_N\times D_{(N+\kappa)}$ rectangular matrix. 
Equivalently, 
we can identify the section of the line bundle over $\C P_N^n$ 
given by $\hat\xi\in C(\kappa,\C^{n+1}_\theta)_N$ 
as $\C P^n_{N+\kappa}$ - $\C P^n_N$ bimodule. 

Note that from this construction, it is apparent that we must impose the 
bound $\kappa+N\geq 0$.

{}From the above construction, we see that there are two pictures of 
the monopole bundle 
$\C P^n_{N+\kappa}$ - $\C P^n_N$ bimodule. Namely, 
the same bimodule can be obtained
from fuzzy 
$\C P^n_{N+\kappa}$ with monopole charge $-\kappa$ 
(assuming $N+\kappa>0$).
Since in noncommutative algebras we have to make a choice of left and right 
multiplication on the module, we can find two equivalent bimodules as
\begin{enumerate}
\item monopole with charge $|\kappa|$ on $\C P^n_N$
\item monopole with charge $-|\kappa|$ on $\C P^n_{N+|\kappa|}$.
\end{enumerate}
Hence
there is a duality between  $\C P_N^n$ with monopole charge $\kappa$
and  $\C P_{N+\kappa}^n$ with monopole charge $-\kappa$.
We see that $\C P^n_N$ and $\C P^n_{N+|\kappa|}$ 
are Morita equivalent, and the scalar field is the equivalence
bimodule
(the inverse of $\hat\xi$ is given by its conjugate).
This is an example showing the relation of Morita equivalence 
and duality of the noncommutative space.

\section{Differential calculus, connection and field strength}

\subsection{Differential forms}

We introduce a basis of one-forms $\theta_a$, $a=1,2,...,n^2+2n$ \`a la Madore
\cite{Madore1991}, which transform in the 
adjoint of $su(n+1)$ and commute with the algebra of functions:
\beq
[\theta_a,f] =0, \qquad \theta_a \theta_b = - \theta_b \theta_a.
\eeq
This defines a space of exterior forms on fuzzy $\C P^n_N$, which we 
denote by $\Omega^*_N:= \Omega^*(\C P^n_N)$. The 
gradation given by the number of anticommuting generators $\theta_a$.
The highest non-vanishing form 
is the $(n^2+2n)$-form 
corresponding to the volume form of $su(n+1)$.

One can also define an exterior derivative 
$d: \Omega^k_N \rightarrow  \Omega^{k+1}_N$ such that $d^2 =0$ and imposing 
the graded Leibniz rule. Its action on the algebra elements 
$f \in \Omega^0_N$ is given by the commutator with a special one-form:
Consider the invariant one-form
\beq
\Theta = \xi_a \theta_a.
\eeq
Then the exterior derivative of a function $f \in \C P^n_N$ is given by
\beq
d f := [\Theta,f] = [\xi_a,f] \theta_a.
\eeq
In particular, we have
\beq
d\xi_b =  [\xi_a,\xi_b] \theta_a = i f_{abc} \theta_a \xi_c.
\label{dxi}
\eeq
The definition of $d$ on higher forms is straightforward, once we find
$d: \Omega^1_{N} \rightarrow \Omega^2_{N}$ such that $d^2(f) =0$.
To find it, we follow the approach of \cite{qfuzzysphere} for the 
$q$-fuzzy sphere. Notice first that  there is a natural bimodule-map 
from one-forms to 2-forms, given by
\beq
\star_1(\theta_a) := \frac i2 f_{abc} \theta_b \theta_c.
\eeq
Then we define
\beqa
d: \; \Omega^1_{N} &\rightarrow& \Omega^2_{N}, \nn\\
                   \a~~ &\mapsto& d\alpha= [\Theta,\a]_+ - \star_1(\a)
\label{d_1}
\eeqa
where $\alpha\in\Omega^1_N$.
One can verify $d^2=0$ in general. To see this, note that
$$
d df = [\Theta,df]_+ -\star_1(df)=0
$$ 
using the following relation:
\beq
\star_1(\Theta) = \Theta^2.
\label{starTheta}
\eeq
This follows from
\beq
\Theta^2 = \Theta \Theta =\half \theta_a \theta_b[\xi_a, \xi_b] 
 = i\half  f_{abc} \theta_a \theta_b\xi_c =: \frac i{\Lambda_N} \eta
\label{theta-2}
\eeq
where $\eta = \frac 12 \theta_a \theta_b f_{abc} x_c$ is the symplectic form.
One can also show that
\beq
d \Theta = \Theta^2, 
\eeq
which implies
\beq
d\eta = 0.
\eeq
Furthermore, 
using group-theoretic arguments\footnote{Using \eq{dxi} we can write
$f_{abc} d x_a d x_b x_c \propto \sum xxx \theta \theta$, which must be
a singlet. Now
$xxx \in V_{(3,3)} \oplus V_{(2,2)}\oplus V_{(1,1)}$, but only $V_{(1,1)}$ can 
be contracted with $\theta \theta$ to give a singlet.} 
it is easy to see that 
\beq
\eta = C \frac 12 f_{abc} d x_a d x_b x_c
\label{c-def}
\eeq
for some numerical constant $C$. This is more easily recognized as 
symplectic form.

In order to extend this calculus to $\C^{n+1}_\theta$,
we can  express the generator $\Theta$ in terms of the
generators $a^j, a^+_i$ 
as above, interpreted as quantizations of the coordinate functions 
$z^j, \obar{z}_i$ on $\C^{n+1}$:
\beq
\Theta = a^+_i {T_a^i}_{j} a^j \theta_a.
\eeq
Then the calculus on $\C P^n_N$ naturally 
induces a calculus on $\C^{n+1}_\theta$.

\paragraph{Relation to the classical case.}

For later use, we want to calculate the constant
$C$ in \eq{c-def} in the classical limit $N \to \infty$.
Consider $\C P^2$ for simplicity. We introduce a 
normalization of the frame by
\beq
\langle \theta_a,\theta_b\rangle = c\d_{ab},
\eeq
where the constant $c$ is determined such that 
the tangential one-forms are properly normalized: using \eq{dxi}, we
have
\beq
\langle d y_a, dy_b\rangle 
 = -f_{ras} f_{ubv} y_s y_v \langle \theta_r, \theta_u\rangle 
 = - c f_{ars} f_{brv} y_s y_v
\eeq
(recall that $y_a$ denotes the classical coordinate functions).
It is sufficient to consider the ``south pole'' of $\C P^2$, where 
$y_a = - \d_{a,8}$ as discussed in section
\ref{sec:coord-class} (setting $r=1$). Then 
\beq 
\langle d y_a, dy_b\rangle  = - c f_{ar8} f_{br8}  
  \;\;\stackrel{!}{=} \;\d_{ab}^{(tang)},
\eeq
which due to the explicit form of $f_{ab8}$
is non-vanishing only for tangential $d y_a$, i.e. 
$a,b \in \{4,5,6,7\}$ as in
\eq{tang-indices}.
This shows that the ``non-tangential'' one-forms on $\C P^2$ have zero
norm, and indeed the correct 4-dimensional calculus on $\C P^2$ 
is recovered from our construction. In the fuzzy case, the additional
one-forms cannot be avoided, however.

The constant $c$ can now be calculated by
summing over the tangential $a,b \in \{4,5,6,7\}$, 
which gives
\beq
4 = \d_{ab}^{(tang)} \langle d y_a, dy_b\rangle = 
\d_{ab} \langle d y_a, dy_b\rangle  = -cf_{ras} f_{rav} y_s y_v 
 = -3 c.
\eeq
Here we used the result that $\langle d y_a, dy_b\rangle=0$ for
non-tangential forms, and
\beq
f_{ras} f_{rav} = 3 \delta_{sv}
\eeq
for $su(3)$. Therefore
\beq
c = - \frac 43.
\eeq
We can now relate $\eta=\frac 12 \theta_a \theta_b f_{abc} y_c$ to
\beq
\eta' = \frac 12 f_{abc} d y_a d y_b y_c.
\eeq
Comparing $\langle \eta,\eta\rangle$ with
 $\langle \eta',\eta'\rangle$, we obtain
\beq
\eta'  = \frac 1{2c} f_{abc} \theta_a \theta_b y_c
 = \frac 1c\eta,
\eeq
therefore
\beq
\eta' = -\frac 3{4} \eta
\label{eta-etaprime-3}
\eeq
for $\C P^2$. For $\C P^n$, the same calculation gives
\beq
\eta' = -\frac{n+1}{2n} \eta.
\label{eta-etaprime-n}
\eeq

\subsection{Canonical Connection and Field Strength}

Once we have a differential calculus, we can define the 
canonical connection $\nabla$ over the projective 
module $\Gamma_\kappa(\C P^n_N)$ 
defined by the projection
$\CP_\kappa$ by 
\be
\CP_\kappa d\xi
\ee
where $\xi\in\Gamma_\kappa(\C P^n_N)$.
The curvature 2-form of this canonical connection is given by
$\CP_\kappa d\CP_\kappa d\CP_\kappa$. 

When the connection is represented 
by the covariant derivative on 
the scalar field $\hat\xi$, we obtain
\be
\nabla\hat\xi \equiv \Bmv^\dagger (\CP_\kappa d\xi)
= (d+\Bmv^\dagger d \Bmv)\hat\xi.
\ee
The gauge field and field strength of the above connection $\nabla$ 
is given by
\be
A=\Bmv^\dagger d \Bmv=\Bmv^\dagger \Theta \Bmv-\Theta
\ee
\beq
F = \Bmv^\dagger \CP_\kappa d\CP_\kappa d\CP_\kappa \Bmv 
  = \Bmv^\dagger\Theta\Bmv \Bmv^\dagger  \Theta \Bmv
    -  \Bmv^\dagger \Theta^2  \Bmv.
\eeq
In order to evaluate this expression, 
we extend the differential calculus
from $\C P^n_N$ to $\C^{n+1}_\theta$ as discussed above, 
postulating that
the $\theta_a$ commute with all $a^i, a_j$ 
(this can also be interpreted
as a calculus on the $U(1)$ 
principal bundle over $\C P^{n}_N$).

Assume first $\kappa>0$.
Then using
\beq
a^j \BN = (\BN+1) a^j, \qquad  \BN a^+_i= a^+_i (\BN+1) 
\eeq
we have 
\beqa
\Bmv^\dagger\Theta\Bmv &=&
       \Bmv^\dagger a^+_i a^j\Bmv\; {T_a^i}_{j}\theta_a\nn\\
 &=&  a^+_i \Bmv^\dagger(\BN+1) \Bmv(\BN+1) a^j\; {T_a^i}_{j}\theta_a\nn\\
&=&  \frac{N-\kappa}{N} a^+_i  a^j\; {T_a^i}_{j}\theta_a =
   \frac{N-\kappa}{N} \Theta
\eeqa
since 
\beq
\Bmv^\dagger(\BN+1)\Bmv(\BN+1) =\frac{\BN+1-\kappa}{\BN+1}
\eeq
by construction.

Similarly,
\beqa
\Bmv^\dagger\Theta^2\Bmv &=&  \Bmv^\dagger (a^+_i {T_a^i}_{j} a^j\;\theta_a)
    (a^+_l {T_b^l}_{k} a^k \;\theta_b) \Bmv \nn\\
 &=& \Bmv^\dagger a^+_i {T_a^i}_{j} {T_b^j}_{k} a^k \; \Bmv\theta_a\theta_b
    +  \Bmv^\dagger a^+_i  a^+_l  a^j a^k \Bmv\;  
    {T_a^i}_{j} {T_b^l}_{k}\theta_a\theta_b 
\eeqa
which, using  
\beq
T_a T_b = \frac 1{2(n+1)} \d_{ab} 
+ \frac 12 (i f_{abc} + d_{abc}) T_c\ ,
\eeq
is
\beqa
\Bmv^\dagger\Theta^2\Bmv 
 &=& \Bmv^\dagger a^+_i 
    (\frac 1{2(n+1)} \d_{ab}\d^i_k + \frac 12 (i {f_{abc}} + {d_{abc}}) 
          {T_c^i}_k) a^k \; \Bmv\theta_a\theta_b  \nn\\
 &&  + a^+_i  a^+_l  \Bmv^\dagger(\BN+2) \Bmv(\BN+2)\; 
     a^j a^k\; {T_a^i}_{j} {T_b^l}_{k}\theta_a\theta_b \nn\\
&=& \frac i2 \Bmv^\dagger a^+_i {f_{abc}} {T_c^i}_k a^k\;\Bmv\theta_a\theta_b
  + \frac{N-\kappa}{N} a^+_i ([a^+_l,  a^j]+ a^ja^+_l) a^k\; 
         {T_a^i}_{j} {T_b^l}_{k}\theta_a\theta_b \nn\\
&=&  \frac i2 a^+_i \Bmv^\dagger(\BN+1)\Bmv(\BN+1)
   f_{abc} {T_c^i}_k a^k\;\theta_a\theta_b \nn\\
 &&  - \frac{N-\kappa}{N} a^+_i  a^k\; 
        {T_a^i}_{j} {T_b^j}_{k}\theta_a\theta_b 
  + \frac{N-\kappa}{N} a^+_i {T_a^i}_{j} a^j a^+_l 
     {T_b^l}_{k} a^k\;  \theta_a\theta_b \nn\\
&=& \frac{N-\kappa}{N} \Theta^2.
\eeqa
Therefore
\bea
F &=& \left( (\frac{\BN-\kappa}{\BN})^2 
- (\frac{\BN-\kappa}{\BN})\right) \Theta^2
={-\kappa (\BN-\kappa)\over \BN^2\Lambda_N}i\eta \cr
&=&{-\kappa N\over (N+\kappa)^2\Lambda_N}i\eta\label{Fplus}
\eea
where we have used that when the field strength $F$ is evaluated
over the scalar field $\hat\xi$, the number operator takes the 
value $\BN=N+\kappa$.
Similarly for $\kappa <0$, we have
\be
F = {|\kappa|(\BN+n+|\kappa|+1)\over(\BN+n+1)^2\Lambda_N}i\eta  
={|\kappa|(N+n+1)\over(N-|\kappa|+n+1)^2\Lambda_N} i \eta.\label{Fminus}
\ee
Thus in the large $N$ limit,
 we obtain for all $\kappa$
\be
F= -\kappa \sqrt{\frac n{2(n+1)}}\; i \eta 
 = \kappa \sqrt{2} \left(\frac{n}{n+1}\right)^{3/2} i \eta'.
\label{F-explicit}
\ee
using \eq{eta-etaprime-n}, up to $+o(1/N)$ corrections.
Hence the field strength is indeed quantized,
and it is a multiple of the symplectic form
in the large $N$ limit. 
In the next section, we verify that  the first Chern number $c_1$ 
is given by $-\kappa$ in the classical limit.

\subsection{Calculation of the first Chern number for $N \to\infty$}
\label{sec:chern-int}

In the classical case, 
we can integrate the symplectic form $\eta'$ over the cycle
$y_1^2+y_2^2+y_3^2 = \frac{n+1}{2n}$ in $\C P^n$ found in 
section \ref{sec:coord-class}. 
Using $f_{abc} = \varepsilon_{abc}$ for $a,b,c \in
\{1,2,3\}$, we have
\beq
\int_{S_R^2} \eta' = 
\int_{S_R^2} \frac 12 \varepsilon_{abc} y_a d y_b d y_c = 4\pi R^3
\eeq
where the sphere has radius $R^2 = \frac{n+1}{2n}$. This shows in
particular that these spheres are indeed non-trivial.
Therefore using \eq{F-explicit}, the first Chern number is
\be
c_1 = \frac i{2\pi}\int_{S_R^2}\; F  = -\kappa
\label{chern-class}
\ee 
in the commutative
limit $N \to \infty$.
This shows that the bundles constructed above
should be interpreted as noncommutative versions of the classical
monopole bundles with Chern number $c_1 =-\kappa$.

\subsection{Discussion on Chern numbers for finite $N$}

In the fuzzy case (i.e. for finite $N$), it is very difficult to give a
satisfactory definition of Chern numbers. One reason for this is the 
lack of a differential calculus with the appropriate dimensions for
finite $N$. However, it is known e.g. from recent investigations of 
fuzzy spheres \cite{GrosseMP0105033} that 
it is still possible to write down suitable integrals in the fuzzy
case, which  in the large $N$ (i.e. the commutative) limit reproduce
the usual Chern numbers, but which are neither topological nor
integer for finite $N$.  We will refer to such 
prescriptions as ``asymptotic'' Chern numbers. 
They are still useful since they produce numbers in the fuzzy case which
reduce to the usual (integer) Chern numbers in the classical limit. 
We illustrate this for fuzzy $\C P^n_N$
by giving a prescription to calculate such an
``asymptotic'' Chern number $c_1$, integrating $\frac i{2\pi} F$
over a suitable ``fuzzy sub-sphere''. 

\subsubsection{Fuzzy sub-spheres}

In the classical case, $c_1$ can be obtained 
by integrating $\frac i{2\pi} F$
over any 2-sphere in $\C P^n$.
Since $\C P^n_N$ is defined in terms of a simple matrix algebra, it
does not admit any non-trivial 
subspaces $\C P^n_N/\cI$ defined by some two-sided
ideal $\cI$. Therefore
in order to compute the first ``asymptotic'' Chern number, we
have to relax the concept of a subspace in the fuzzy case. 
A natural way to do this in our context is the following:

Note that for any given root $\a$ of $su(n+1)$,
$V_N \cong \cF_N$ decomposes into a direct sum of irreps 
of $su(2)_\a \subset su(n+1)$. 
Now fix a root $\a$. Then there is precisely one such irrep denoted by
$H^{\a,N}$ which has maximal dimension $N+1$
(note that the weights of $V_N$ form a simplex in weight lattice of
size $N+1$).
The other irreps have smaller dimension
$M+1 \leq N+1$, denoted by $H^{\a,M}$ (we omit additional
labels for simplicity). Let $P^{\a,M}: V_N \to H^{\a,M}$ 
be the projector on $H^{\a,M}$.
Now we can define maps
\bea
\C P^n_N \cong End(V_N) &\to&  End(H^{\a,M)})  \nn\\
  f &\mapsto& \hat f := P^{\a,M}f P^{\a,M}.
\eea
In principle, 
each 
\be
S^2_{\a,M} := End(H^{\a,M})
\label{s2Ma}
\ee
could be considered as a fuzzy sphere,
but not necessarily as
sub-spheres of $\C P^n_N$. However, we 
shall explain below  
that the maximal $S^2_{\a,N}$
can be considered as an ``asymptotic sub-sphere'' of  $\C P^n_N$,
in the sense that it becomes the algebra of functions on a sub-sphere
of $\C P^n$ in the large $N$ limit. 
This sub-sphere in fact coincides 
with the non-trivial 2-cycles found in 
section \ref{sec:coord-class} for suitable choice of  the root $\alpha$. 
To see this, consider the corresponding projected coordinate generators
\be
\hat x_a =  P^{\a,M} x_a P^{\a,M}
\ee
obtained from the fuzzy coordinate functions of $\C P^n_N$.
If $S^2_{\a,M}$ is to be interpreted as sub-sphere 
of $\C P^n_N$, then $\hat x_a$ should be
interpreted as restriction (or pull-back) of the coordinate function
$x_a$ of $\C P^n_N$ to $S^2_{\a,M}$. However, it is easy to see 
that\footnote{this inequality is to be understood in the operator-norm}
\be
g_{ab} \hat x_a \hat x_b < 1
\label{constraint-hat}
\ee
for finite $N$, which is in contrast to the constraint 
$g_{ab} x_a  x_b = 1$  \eq{defz2-n} of $\C P^n_N$.
The reason is that the (rescaled) quadratic Casimir of $su(n+1)$,
which can be written as
\be
g_{ab} x_a  x_b = \sum_\b (\frac 12 x_\b^{+} x_\b^{-} + \frac 12  x_\b^{-} x_\b^{+}) +
\sum_i H_i^2 = 1,
\label{casimir-uncon}
\ee 
where $ \sum_\b$ goes over all positive roots of $su(n+1)$, and
$H_i$ are the (suitably rescaled) Cartan generators. 
Now the restricted generators are related to the unrestricted ones as
follows:
\beqa
\hat x_\a^{\pm} &=&  x_\a^{\pm}, \quad \hat H_i = H_i \nn\\
\hat x_\b^{\pm} &=&  0 \quad \mbox{for} \;\; \b \neq \a, 
\eeqa
because $x_\b^{\pm}$ does not preserve any $H^{\a,M}$ for $\b\neq \a$.
This implies \eq{constraint-hat}, 
since $(\frac 12 x_\b^{+} x_\b^{-} + \frac 12  x_\b^{-} x_\b^{+})$ 
is positive definite.
This reflects  the
fact that $\C P^n_N$ does not admit any (strict) sub-spaces. 
However, we can consider ``asymptotic sub-spaces'' by relaxing 
the constraint $g_{ab} x_a x_b = 1$ and 
allow for ``quantum corrections'' of order $1/N$, 
requiring only
\be
g_{ab} \hat x_a \hat x_b = 1 - O(1/N).
\label{constraint-hat-mod}
\ee
This holds indeed for $H^{\a,N}$, but in general not for $H^{\a,M}$ 
with $M < N$.
To see this, recall that the rising- and lowering 
operators act as 
$\xi_\b^+ v_k = \sqrt{(M-k) k}\; v_{k+1}$ where $\{v_k\}$ is the
normalized weight basis of 
the $su(2)_\b$ irrep $H^{\b,M}$, and similar for $\xi_\b^-$. 
Including the scale
factor $\L_N = O(1/N)$, this implies
\be
x_\b^\pm = O(1/\sqrt{N})
\label{statementX}
\ee 
if acting on $H^{\a,N}$ 
(recall that the weights of $V_N$ form a simplex in weight lattice of
size $N$, and $H^{\a,N}$ forms an edge of this simplex), 
while it is not true in the bulk of $V_N$ i.e. on general $H^{\a,M}$.
Therefore we can consider
$S^2_{\a,N}$ as ``asymptotic subspace'' of $\C P^n_N$,
which in fact reduces for $N \to \infty$ 
to the non-trivial 2-cycles found in 
section \ref{sec:coord-class}. 
This will be used in the calculation of asymptotic first Chern numbers 
below.

\subsubsection{Asymptotic first Chern number}

To compute the Chern number $c_1$, we have to integrate ${i\over2\pi}F$
over a 2-cycle in $\C P^n_N$. According to the above discussion, we 
give a definition in terms of an integral 
over a specific fuzzy 
$S^2_{\a,N} \cong \C P^1_N$ which is an asymptotic subspace
of $\C P^n_N$ in the above sense, and then perform the integration.

In order to specify the $S^2_{\a,N}$, we first choose a 
Hilbert space $H^{\alpha, N}$ with maximal dimension,
denoted by $\CF_S$. It can be defined as
\be
\CF_S=\{\Ket{\psi}\in\cF_N;\;\; a_i\Ket{\psi}=0 \mbox{~for~ }i=3,...n+1\mbox{~and~}\BN\Ket{\psi}=N\Ket{\psi}\}\ .
\ee 
The space $\CF_S$ has the same dimension as the Hilbert 
space of fuzzy $\C P^1_N$. The generators associated 
with the $su(2)$ rotations of this fuzzy $\C P^1$ 
are 
\be
T_{m}= {1\over 2}
\left(\begin{array}{ccc}
\sigma^m&\cdots&0\cr\vdots&\ddots&\vdots\cr0&\cdots&0
\end{array}\right).
\ee
The corresponding coordinates are linear combinations 
of coodinate operators of $\C P^n_N$, and 
we denote them by $\hat x_m=\Lambda_N a^+_iT^m_{ij}a_j$, in 
agreement with 
the notation of the previous section.
The radius of fuzzy $\C P^1$ is defined by these 
coordinates $\hat x_m$ as
\be
\sum_{m=1}^3 (\hat x_m)^2={1\over4}\Lambda_N^2 N(N+2)=R_N^2\ .
\ee
Representing the algebra generated by $\hat x_m$ on $\CF_S$, we obtain 
precisely the matrix algebra of fuzzy $\C P^1_N$. Therefore we can use the 
standard results of the integration over $\C P^1_N$. 
We introduce the volume element of this fuzzy $\C P^1_N$, 
$$
\omega={1\over 2R_N}\epsilon_{mnp}\hat x_md\hat x_nd\hat x_p
\quad \in \Omega^2
$$ 
which is an invariant 2-form and
agrees with the volume element of the 
sphere with radius $R=\sqrt{n+1\over2n}$ in the commutative limit 
$N\rightarrow\infty$.
The integration $\int: \Omega^2\rightarrow \C$ over this
fuzzy $\complex P^1_N$
is then defined by \cite{GrosseMP0105033} 
\be
{1\over2}\int_{\C P^1_N}  f \epsilon_{mnp}\hat x_md\hat x_nd\hat x_p
\equiv 4\pi R_N^3 Tr_{\CF_S}\{f\}\ ,
\label{CP1-int}
\ee 
where $Tr_{\CF_S}$ denotes the trace over  $\CF_S$ 
normalized such that $Tr_{\CF_S}\{1\}=1$.
Using the commutation relations of $\hat x_m$ and the 
definition of the derivatives 
defined in the previous section, we obtain 
\bea
\omega
&=&R_N({1\over\BN(\BN+2)}-\half) \epsilon_{mnp}\hat x_m\hat\theta_n\hat\theta_p.
\eea
where the $\hat\theta_m$, m=1,2,3 is the one form over $\C P^1_N$. 

The Chern character $c_1$ is defined by using the 
projection operators $\CP_\kappa$. To evaluate it we  
first take the  $\C P^n_N$ -valued trace 
$Tr_\kappa\{\cdot\}$
over $End(\Gamma_\kappa(\C P^n_N))$, and then
integrate over $\C P^1_N$: 
\be
c_1={i\over 2\pi}\int_{\C P^1_N} 
Tr_\kappa \{\CP_\kappa d\CP_\kappa d\CP_\kappa\}=
{i\over 2\pi}\int_{\C P^1_N}  \sum_\mu 
v_\mu F_\kappa v^\dagger_\mu\label{chern1}
\ee
where $F_\kappa$ is the field strength 
$F$ given either in eq.(\ref{Fplus}) 
or eq.(\ref{Fminus}), depending on the value of $\kappa$. 

Using the definition \eq{basevector+}, \eq{basevector-} 
of $v_\mu$ we can perform the summation over $\mu$ and find 
the following expression for the integrand:

For $\kappa>0$:
\be
\sum_\mu v_\mu F_\kappa v^\dagger_\mu
={(\BN+n+\kappa+1)\cdots (\BN+n+2) \over(\BN+1)\cdots (\BN+\kappa)}{-i\kappa\BN\over 2(\BN+\kappa)^2\Lambda_N}f_{abc}\,x_a \theta_b\theta_c
\ee
For $\kappa<0$:
\be
\sum_\mu v_\mu F_\kappa v^\dagger_\mu
={-i\kappa(\BN-|\kappa|)\cdots(\BN-1)\over(\BN+n-|\kappa|+1)\cdots (\BN+n)} {(\BN+n+1)\over 2(\BN-|\kappa|+n+1)^2\Lambda_N}f_{abc}\,x_a \theta_b\theta_c
\ee
Theses are 
$2$-forms over $\C P^n_N$. In order to integrate 
them over the fuzzy sub-space $\C P^1_N$,
we should also pull-back these $2$-forms to $\C P^1_N$.
To do this we split the coordinates into
the two orthogonal sets, 
$(x_m, x_{m^\perp})$, where $x_m$ corresponds to the 
$SU(2)_\alpha$.
Correspondingly, we split the one-forms into
$(\theta_m,\theta_{m^\perp})$. 
This means that
$\theta_m$ is the dual of 
$\partial_m={1\over \Lambda_N} ad_{x_m}$,
analogously to the commutative case.
Now we define the pull-back by projecting out 
$\theta_{m^\perp}$ and identify $\hat\theta_m$ with
$\theta_m$.

Since $\C P^1_N$ corresponds to a $su(2)$
subalgebra of $su(n+1)$, this implies that the pull-back of 
$f_{abc}\, x_a \theta_b\theta_c$ is 
$\epsilon_{mnp}\hat x_m \hat\theta_n\hat\theta_p$. 
This can now be integrated over fuzzy $\C P^1_N$ with radius $R_N$
according to \eq{CP1-int},
and  we get from (\ref{chern1})
for $\kappa>0$
\bea
c_1
&=&{i\over 2\pi}{(N+n+\kappa+1)\cdots (N+n+2) \over(N+1)\cdots (N+\kappa)}{-i\kappa N\over 2(N+\kappa)^2\Lambda_N}\int_{\C P^1_N}  
\epsilon_{mnp}(\hat x_m \hat\theta_n\hat\theta_p)\cr
&=&-\kappa {(N+n+\kappa+1)\cdots (N+n+2) \over(N+1)\cdots (N+\kappa)}{ N\over (N+\kappa)^2}{\sqrt{N(N+2)}\over (1-{2\over N(N+2)})}
\eea
For large $N$ this yields
\be
c_1=-\kappa+{1\over N}(\kappa(n-1)+1)+\cdots
\ee

In the same way we get for $\kappa<0$:
\be
c_1
=-\kappa
{(N-|\kappa|)\cdots(N-1)\over(N+n-|\kappa|+1)\cdots (N+n)} {(N+n+1)\over(N-|\kappa|+n+1)^2}{\sqrt{N(N+2)}\over (1- {2\over N(N+2)})}
\ee
The expansion with respect to ${1\over N}$ is 
\be
c_1=-\kappa(1+{1\over N}(-\kappa (n-1)-n)+\cdots)
\ee

\section{Conclusion}

In this paper, we investigated the definition of 
fuzzy complex projective space $\C P^n_N$ from two different points of view,
and constructed the nontrivial $U(1)$ bundles over those spaces. 
The corresponding Chern classes are calculated. 

The first approach is to consider $\C P^n$ as (co)adjoint orbit, given by 
$(n+1) \times (n+1)$ matrices $Y$  which satisfy a certain characteristic
equation. The quantization of the function algebra
is given by a simple matrix algebra
$Mat(D_N,\C)$, more precisely $End_\C(V_N)$ for certain
irreducible \reps $V_N$ of $su(n+1)$. The appropriate \reps  $V_N$
are determined using harmonic analysis. This leads to an 
algebra-valued $(n+1) \times (n+1)$ matrix $X$, whose
characteristic equation gives the explicit relations satisfied
by the fuzzy coordinate functions.

The second approach uses the generalized Hopf fibration
$U(1)\rightarrow S^{2n+1}\rightarrow \C P^n$.
Again a characteristic equation is derived for 
a certain operator-valued 
matrix, which coincides with the first approach
when we specify the Fock space representation $\CF_N \cong V_N$.
The second construction is very useful to define the projective modules.

We then construct the projective modules by giving the projection
operator in terms of a normalized vector, following the approach for 
monopoles on $S^2$. 
We find nontrivial projective modules
of $\C P^n_N$ labeled by an integer $\kappa$, which are
interpreted as fuzzy version of
the monopole bundles on $\C P^n$ with monopole number $\kappa$. 
Using a suitable differential calculus, we then calculate
the field strength over the monopole bundle, or equivalently the 
first Chern class. 
We verify explicitly that the usual Chern number $c_1$ is recovered in 
the commutative limit.

Finally let us recall that fuzzy spaces arise naturally in string theory, 
for example as D-branes on group manifolds or as solutions of the IKKT 
matrix model. In both cases one expects that the low-energy 
effective action should be given by an induced gauge theory.
These gauge theories typically have degrees of freedom which are
not tangential as in conventional field theories. As we
have seen, such degrees of freedom do arise naturally in the
differential calculus on fuzzy spaces.

There are several open questions which deserve further
investigations. There exists a somewhat different (but related) 
formulation of monopoles as solutions of a 
matrix model on the fuzzy sphere \cite{fuzzyYM}; the  
extension of this construction 
to fuzzy $\C P^2$ will be presented elsewhere \cite{HG}. 
Furthermore, fermions have been discussed
extensively on the fuzzy sphere, but for $\C P^n$ no fully
satisfactory formulation is available. 
Of course, a generalization to instantons would also be desirable.

\medskip\medskip
\noindent{\bf \large Acknowledgement} 
The authors would like to thank J. Madore, G. Landi and B. Jur\v co for 
discussions. S.W. would also like to thank 
J. Wess and B. Jur\v co for their hospitality at M\"unchen University, 
where part of this work was performed.
H.S. also thanks H. Grosse for discussions, and
the organizers of the 
'International Workshop on Noncommutative Geometry and Physics' 
at Keio University 2004 
for an invitation, where part of this work was performed.
This research is partly supported by Grant-in-Aid for Scientific 
Research from the Ministry of Education, Culture, 
Sports, Science and Technology, Japan No. 13640256.

\section{Appendix: characteristic equation}

\subsection{$su(3)$}

Consider the \rep
\beq
V:= V_{N \Lambda_1} \tens V_{\Lambda_1} = V_{(N+1)\Lambda_1} \oplus V_{(N+1)\Lambda_1-\a_1}
\label{V-decomp}
\eeq
of $su(3)$. The  operator $X = \sum_a \xi_a \la_a$ 
is an intertwiner on $V$, 
where $\la_a  =2 \pi_{\Lambda_{(1)}}(T_a)$ and
$\xi_a = \pi_{N\Lambda_{(1)}}(T_a)$   and 
$T_a$ are the generators of $su(3)$
which satisfy $[T_a,T_b] = i f_{abc} T_c$.
$X$ can be related to the quadratic Casimir of $su(3)$,
\beq
C_2 = 2\sum_a T_a T_a 
\eeq
as follows:
\beq
X = \sum_a \xi_a \la_a = 2\sum_a \pi_{N \Lambda_1}(T_a) \pi_{\Lambda_1}(T_a) 
= \frac 12 (\pi_{V}(C_2) - \pi_{N \Lambda_1}(C_2) - \pi_{\Lambda_1}(C_2)).
\eeq
Recall that the eigenvalues of the quadratic Casimirs 
on the highest weight \rep $V_{\Lambda}$ are given by
\beq
C_2(\Lambda) = (\Lambda, \Lambda+2\rho) 
\eeq
where $\rho = \sum \Lambda_i$ 
is the Weyl vector of $su(3)$, and $(,)$ denotes the Killing form.
It follows that the eigenvalue of $X$ 
on the components $V_{N \Lambda_1+\nu}$  in \eq{V-decomp} is
\beq
X = \frac 12 (C_2(N\Lambda_1 + \nu) - C_2(N\Lambda_1) - C_2(\Lambda_1))
  = (\nu, N\Lambda_1) + (\nu-\Lambda_1,\rho)  
\eeq
for $\nu = \Lambda_1$ resp. $\nu =  \Lambda_1-\a_1$.
Using the inner products of the fundamental weights
\beq
(\Lambda_1,\Lambda_1) = \frac 2{3} = (\Lambda_2,\Lambda_2), \quad (\Lambda_1,\Lambda_2) = \frac 1{3}, 
 \quad (\Lambda_2,\rho) = 1 = (\Lambda_1,\rho) 
\eeq
for $su(3)$, we find the eigenvalues of $X$ as 
$(\frac{2N}{3},-\frac {N}{3} -1)$,
hence the characteristic equation of $X$ is 
\beq
(X-\frac{2N}{3})(X+\frac{N}{3}+1) =0.
\label{char-X-n}
\eeq

\subsection{$su(n+1)$}

For $\C P^{n}$, consider the \rep 
\beq
V:= V_{N \Lambda_1} \tens V_{\Lambda_1} 
       = V_{(N+1)\Lambda_1} \oplus V_{(N+1)\Lambda_1-\a_1}
\label{V-decomp-n}
\eeq
of $su(n+1)$.  The  operator $X = \sum_a \xi_a \la_a$ 
is an intertwiner on $V$, 
where $\pi_{\Lambda_1}(T_a) = \frac 12 \la_a$ and
$\xi_a = \pi_{N\Lambda_1}(T_a)$  and 
$T_a$ are the generators of $su(n+1)$
which satisfy $[T_a,T_b] = i f_{abc} T_c$.
$X$ can again be related to the quadratic Casimir 
$C_2 = 2\sum_a T_a T_a$
of $su(n+1)$ as follows:
\beq
X = \sum_a \xi_a \la_a 
=  \frac 12 (\pi_{V}(C_2) - \pi_{N \Lambda_1}(C_2) - \pi_{\Lambda_1}(C_2)).
\eeq
Hence the eigenvalue of $X$ 
on the component $V_{N \Lambda_1+\nu}$  in \eq{V-decomp} is
\beq
X = (\nu, N\Lambda_1) + (\nu-\Lambda_1,\rho)
\eeq
for $\nu = \Lambda_1$ resp. $\nu = \Lambda_1-\a_1$.
The eigenvalues are then $(\frac{nN}{n+1},-\frac {N}{n+1} -1)$,
and the characteristic equation of $X$ is given by
\beq
(X-\frac{nN}{n+1})(X+\frac{N}{n+1}+1) =0.
\label{char-X}
\eeq

\eject

\end{document}